\documentclass[]{interact}

\usepackage{epstopdf}
\usepackage[caption=false]{subfig}

\usepackage[numbers,sort&compress]{natbib}
\bibpunct[, ]{[}{]}{,}{n}{,}{,}
\renewcommand\bibfont{\fontsize{10}{12}\selectfont}

\theoremstyle{plain}

\theoremstyle{definition}

\theoremstyle{remark}

\begin{document}

\articletype{}

\title{Stochastic Model of SIR Epidemic Modelling}

\author{
\name{Kurnia Susvitasari\textsuperscript{1,2}\thanks{K.~Susvitasari. Email: susvitasari@sci.ui.ac.id}}
\affil{\textsuperscript{1} School of Mathematical Science, University of Nottingham, UK; \textsuperscript{2} Department of Mathematics, University of Indonesia, Indonesia}
}

\maketitle

\begin{abstract}
	Threshold theorem is probably the most important development of mathematical
	epidemic modelling. Unfortunately, some models may not behave according to
	the threshold. In this paper, we will focus on the final outcome of SIR model with
	demography. The behaviour of the model approached by deteministic and 
	stochastic models will be introduced, mainly using simulations. Furthermore,
	we will also investigate the dynamic of susceptibles in population in absence of 
	infective. We have successfully showed that both deterministic and stochastic models 
	performed similar results when $R_0 \leq 1$. That is, the disease-free stage in the 
	epidemic. But when $R_0 > 1$, the deterministic and stochastic approaches had 
	different interpretations.
\end{abstract}

\begin{keywords}
	SIR with demography, stochastic model, forward Kolmogorov
\end{keywords}

\section{Introduction}
	
	In epidemic modelling, the deterministic and stochastic approximations were use 
	to model the behaviour, especially to know the final outcome of the epidemic.
	Both models were important in any sense to describe this behaviour process.
	The deterministic model, in fact, is also an approximation of the stochastic model 
	when the population size is sufficiently large.
	
	Threshold theorem is the most important occurence in the development of the 
	mathematical theory of epidemic. The threshold behaviour is usually expressed
	in term of epidemic basic reproduction number, $R_0$. This quantity is usually
	defined as the expected number of contacts made by a typical infectious individual
	to any susceptibles in the population. It is important to note that in general epidemic 
	modelling, both stochastic and deterministic models have similar threshold value, which 
	is attained at $R_0 = 1$. It then turns out that the models identify two parameter regions, 
	$R_0 \leq 1$ and $R_0 > 1$, with qualitatively different behaviour. In the deterministic 
	model, minor epidemic always occurs if $R_0 \leq 1$ with probability one, otherwise the 
	major epidemic occurs also with probability one. So, the deterministic model simply 
	identifies the process behaviour according to the value of $R_0$. But in the stochastic 
	model, if we let the size of population be sufficiently large, minor epidemic definitely 
	occurs when $R_0 \leq 1$, whilst when $R_0 > 1$ major epidemic occurs with probability 
	$p \in (0,1)$. It means that there is non-zero probability that the epidemic will die out 
	even though $R_0 > 1$. This is an important difference between the two approaches.
	
	In this paper, we will focus on the stochastic behaviour of the SIR epidemic model.
	Unlike SIS, we model SIR using the 2--dimensional Markov chain. We will also
	consider an SIR model with demography. Tabel \ref{tab:sir.rate} represents the transition 
	rates of the SIR model in this paper.
	
	\begin{table} 
		\caption{Transition rates of SIR model with demography} \label{tab:sir.rate}
		\begin{center}
			\begin{tabular}{|| c c c ||} 
				\hline \hline
				Description				&	Transitions								&	Rates  \\
				\hline
				birth of susceptibles 	&	$(S,I,R) \rightarrow	(S+1,I,R)$	&	$\mu n$	 \\
				death of susceptibles	&	$(S,I,R) \rightarrow (S-1,I,R)$	&	$\mu S$ \\
				death of infectives		&	$(S,I,R) \rightarrow (S,I-1,R)$	&	$\mu I$ \\
				death of removed		&	$(S,I,R) \rightarrow (S,I,R-1)$	&	$\mu R$ \\
				infection 					&	$(S,I,R) \rightarrow (S-1,I+1,R)$ & $\lambda SI/n$ \\
				removal						&	$(S,I,R) \rightarrow (S,I-1,R+1)$ & $\gamma I$\\
				\hline \hline
			\end{tabular}
		\end{center}
	\end{table}
	
	We define a fixed birth rate of susceptibles, $\mu n$, and death rates of susceptible,
	infectious, and immuned individuals as $\mu S, \mu I, \mu R$. In contrast with 
	model without demography, according to N\aa sell \cite{nasell}, the SIR model with demography admits an almost 
	stationary behaviour, corresponding to endemic infectious. Furthermore, we will also 
	investigate the dynamic of susceptibles in population in absence of infective.

\section{The SIR Model with Demography}

	In the model with demography, the population size is not closed, but actually
	depends upon the size of $S, I, \text{ and } R$ at time $t$. But in this case, since
	the birth rate of susceptibles is constant $\mu n$, the dynamical number of 
	population due to the death rates will hardly affect the population size. Therefore,
	it is enough to only track the size of susceptibles and infectious individuals only.
		
	\begin{figure}
		\centering
		\subfloat[Susceptibles vs infectives plot.]{%
		\resizebox*{5cm}{!}{\includegraphics{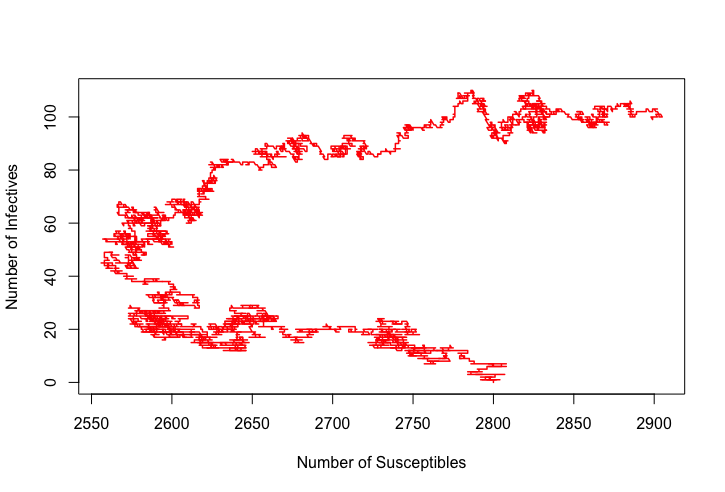}}}\hspace{5pt}
			\label{fig:sir.suscep.vs.inf.r=1}
		\subfloat[Infectives vs time plot.]{%
		\resizebox*{5cm}{!}{ \includegraphics{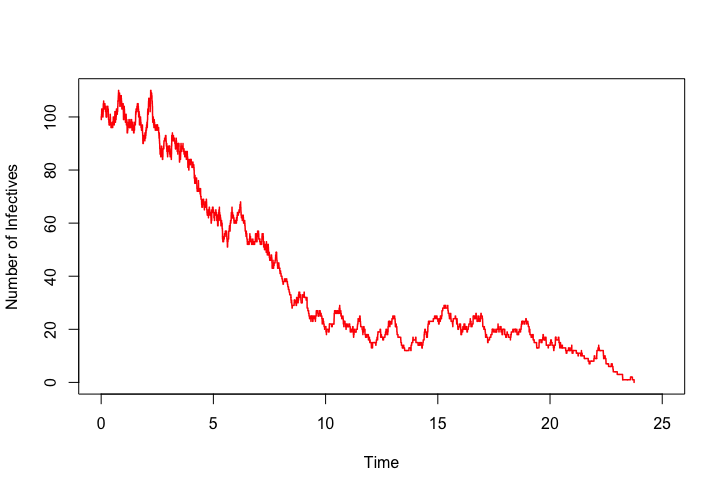}}}
			\label{fig:sir.inf.vs.time.r=1}
		\caption{SIR with demography with $(R_0,\mu,\gamma) = (1,0.1,1)$ and 	
			$(S(0),I(0))=(2900,100)$}
		\label{fig:sir.r=1}
	\end{figure}
	
	\begin{figure}
		\centering
		\subfloat[$R_0 = 3$.]{%
		\resizebox*{5cm}{!}{\includegraphics{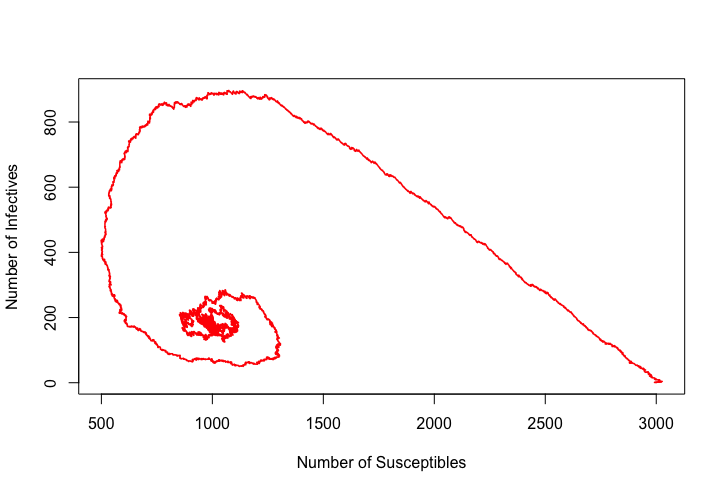}}}\hspace{5pt}
			 \label{fig:sir.suscep.vs.inf.r=3}
		\subfloat[$R_0 = 4,2,1.5$.]{%
		\resizebox*{7cm}{!}{\includegraphics{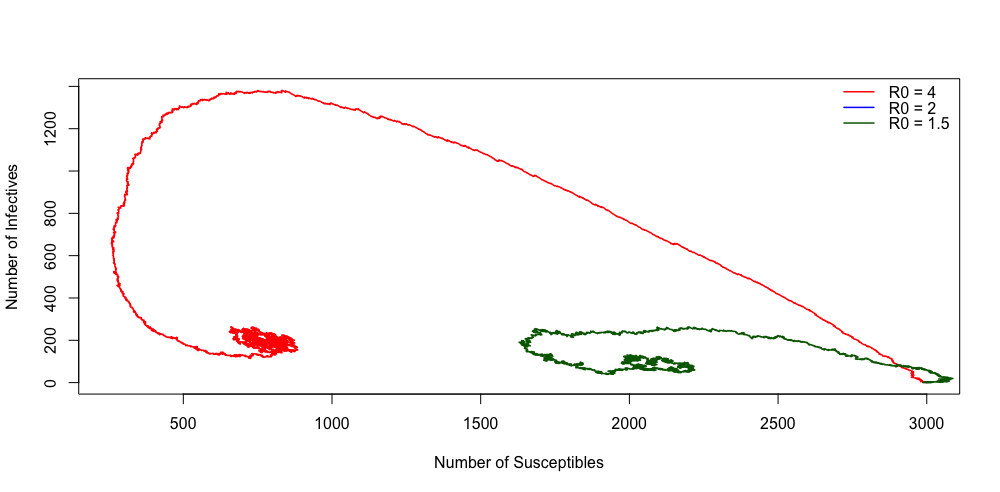}}}
			\label{fig:sir.3plots}
		\caption{SIR with demography with $(\mu,\gamma) = (0.1,1)$ and 
			$(S(0),I(0))=(2999,1)$}
	\end{figure}
		
	Similar to SIS model, SIR model also has threshold at $R_0 = 1$. In Figure 
	\ref{fig:sir.r=1}, the epidemic dies out even though we assign quite a large 
	number of initial infectious individuals, but in Figure \ref{fig:sir.suscep.vs.inf.r=3}a,
	the epidemic takes off and survives in the first cycle. Another interesting
	fact is shown in Figure \ref{fig:sir.3plots}b, one sample dies out quickly even 
	though $R_0 = 2 > 1$. The normal scenario is that epidemic will die out
	if and only if $R_0 \leq 1$, and take off otherwise due to eq.(\ref{eq:sir.deter.suscep.rate})
	 and (\ref{eq:sir.deter.inf.rate}).
	But an anomaly occurs, which cannot explained by the deteministic model. 
	There is a non--zero probability that the SIR will die out even though $R_0 > 1$.

\subsection{The Deterministic Model of SIR with Demography}

	We define the SIR model using 2--dimensional Markov chain process. Suppose
	the process $\lbrace (X_n(t), Y_n(t)) : t \geq 0 \rbrace$ be the number of susceptibles 
	and infectious individuals at time $t$, given inially size $n$ population.
	
	The transition schemes of SIR model with demography is as follows
	\begin{subequations}
		\begin{align}
			\frac{d X_n(t)}{dt} 
			&= \mu \left( n - X_n(t) \right) - \frac{\lambda}{n}X_n(t) Y_n(t) 		
			\label{eq:sir.deter.suscep.rate} \\
			\frac{d Y_n(t)}{dt} 
			&= \frac{\lambda}{n}X_n(t) Y_n(t) - (\mu+\gamma) Y_n(t). 
				\label{eq:sir.deter.inf.rate}
		\end{align}
	\end{subequations}
	
	Suppose that $x(t) = \frac{\displaystyle X_n(t)}{\displaystyle n}$ and $y(t) = 
	\frac{\displaystyle Y_n(t)}{\displaystyle n}$ denote the proportion of number of 
	susceptibles and infectives at time $t$ respectively. Then, we can scale eq.
	(\ref{eq:sir.deter.suscep.rate}) and (\ref{eq:sir.deter.inf.rate}) as follows	
	\begin{subequations}
		\begin{align}
			\frac{d x(t)}{dt} 
			&= \mu \left( 1 - x(t) \right) - \lambda x(t) y(t) \label{eq:sir.prop.suscep} \\
			\frac{d y(t)}{dt} 
			&= \lambda x(t) y(t) - (\mu + \gamma) y(t). \label{eq:sir.prop.inf}
		\end{align}
	\end{subequations}
	
\begin{figure}[ht]
		\centering
		\resizebox*{10cm}{!}{\includegraphics{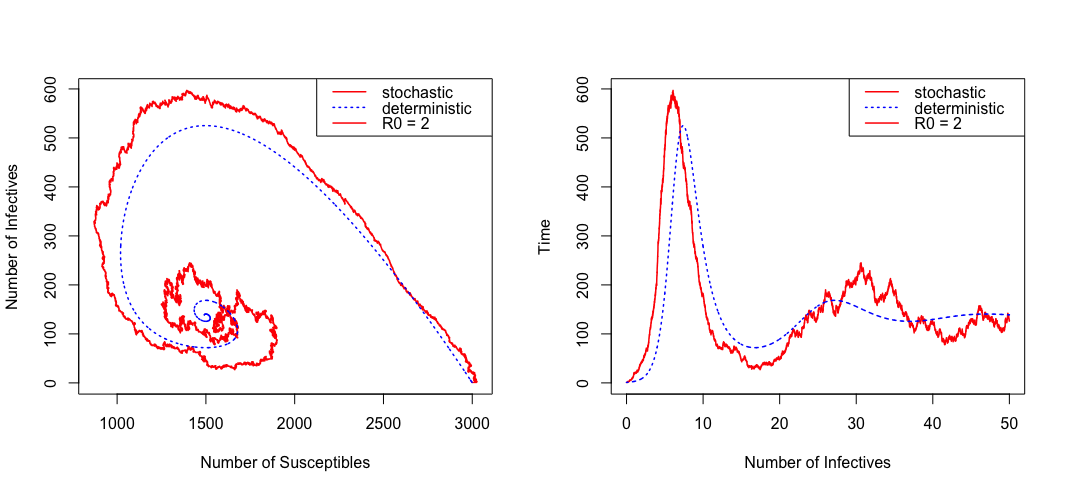}}
		\caption{SIR epidemic with parameter $(R_0,\mu,\gamma) = (2,0.1,1)$} 
		\label{fig:sir.r2}
	\end{figure}

	Now we are interested to find the equilibrium points of the eq. (\ref{eq:sir.prop.suscep})
	and (\ref{eq:sir.prop.inf}). Setting $\displaystyle \frac{d x(t)}{dt} = 0$ and 
	$\displaystyle \frac{d y(t)}{dt} =0$ yields two equilibrium points. Suppose that
	$R_0 = \displaystyle \frac{\lambda}{\mu + \gamma}$, equilibrium point $(1,0)$
	is attained if and only if $R_0 \leq 1$, and $\left( R_0^{-1}, \displaystyle \frac{\mu}
	{\mu + \gamma} - \frac{\mu}{\lambda} \right)$ if and only if $R_0 > 1$. Using 
	Theorem 4.1 in \cite{susvitasari}, the stochastoc process will mimic the behaviour of the
	deterministic process as $t \rightarrow \infty$ (see Figure \ref{fig:sir.r2}). 
	
	\begin{figure}[ht]
		\centering
		\resizebox*{10cm}{!}{\includegraphics{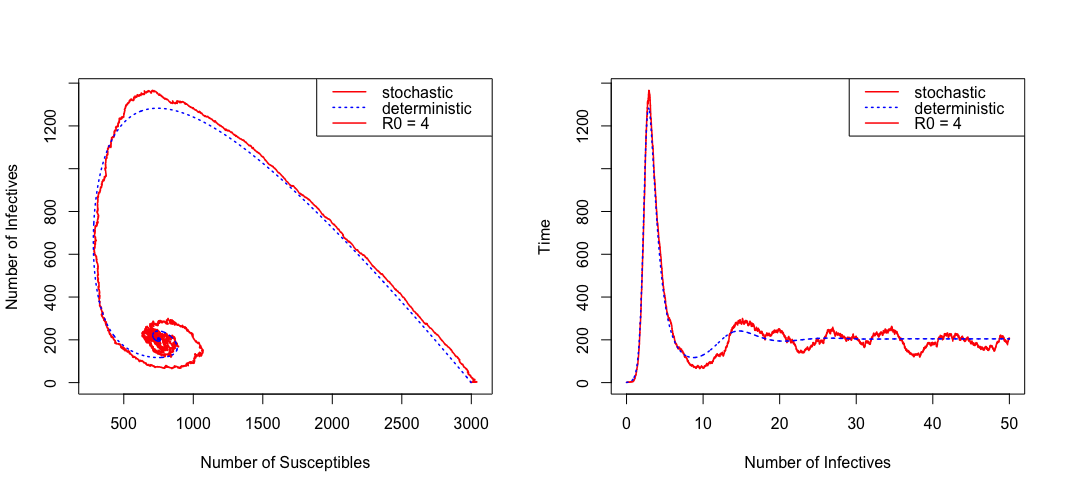}}
		\caption{SIR epidemic with parameter $(R_0,\mu,\gamma) = (4,0.1,1)$} 
		\label{fig:sir.r4}	
	\end{figure}
	
	\begin{figure}[ht]
		\centering
		\resizebox*{10cm}{!}{\includegraphics{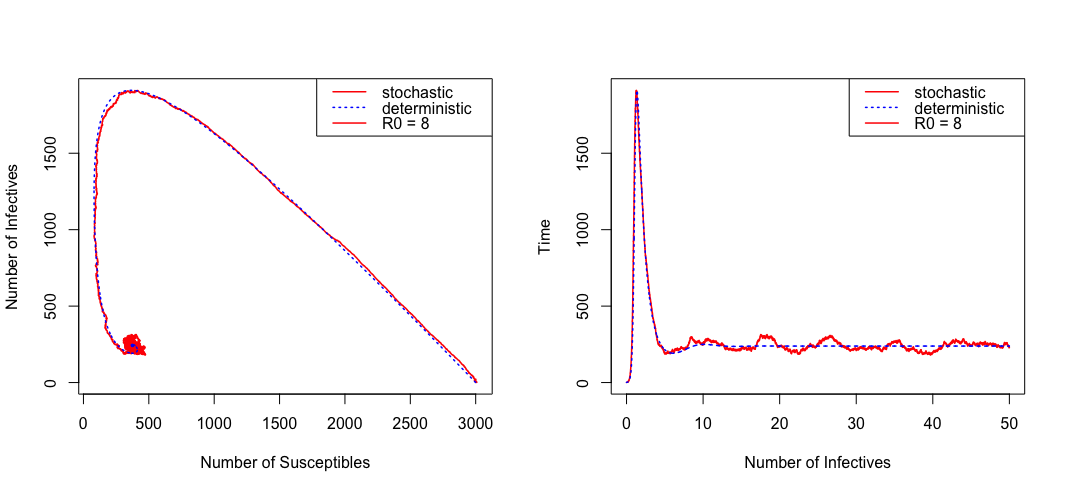}} 
		\caption{SIR epidemic with parameter $(R_0,\mu,\gamma) = (8,0.1,1)$} 
		\label{fig:sir.r8}
	\end{figure}

\subsection{The Stichastic Model of SIR with Demography}
	
	Suppose that $p_{sir}(t) = P \lbrace S(t) = s, I(t) = i, R(t) = r \rbrace$. Using the 
	transition rates in Table \ref{tab:sir.rate}, we can construct forward Kolmogorov
	equation as follows
	\begin{align}
		p_{sir}'(t) &= p_{s-1,i,r}(t) \cdot (\mu n) + p_{s+1,i,r}(t) \cdot (\mu (s+1))+ 
			p_{s,i+1,r}(t) \cdot \nonumber \\
		& \quad (\mu (i+1)) + p_{s,i,r+1}(t) \cdot (\mu (r+1)) + p_{s+1,i-1,r}(t) \cdot 
			\nonumber \\
		& \quad  \left( \frac{\lambda}{n} (s+1)(i-1) \right) +  p_{s,i+1,r-1}(t) \cdot 
			(\gamma (i+1)) - p_{sir}(t) \cdot \nonumber \\
		& \quad \left( \mu (n + s + i + r) + \frac{\lambda}{n} si + \gamma i \right). 		
		\label{eq:sir.forward.final}
	\end{align}
	
	Now, consider Figures \ref{fig:sir.r2}, \ref{fig:sir.r4}, and \ref{fig:sir.r8}. The SIR 
	process certainly has similar behaviour to SIS. In Figure \ref{fig:sir.r2}, we can 
	see a certain deviance between the deterministic and the stochastic models, 
	but in Figures \ref{fig:sir.r4} and \ref{fig:sir.r8} the stochastic model  seems to 
	settle down in equilibrium faster than in Figure \ref{fig:sir.r2} and the deterministic 
	model also approximates nicely. This equilibrium is reached at point $\left(
	 R_0^{-1}, \frac{\displaystyle \mu
	 }{\displaystyle \mu + \gamma} - \frac{\displaystyle \mu }{\displaystyle \lambda} \right)$. 
	 By changing $\mu$, we can see that the threshold theorem is still 
	 applied but the equilibrium point is shifted. 

	Another interesting fact about SIR epidemic modelling with demography is 
	how the dynamic of susceptible size in the absence of infective. Recall equation 
	(\ref{eq:sir.deter.suscep.rate}). In the absence of infective, the proportion of 
	susceptibles in the population is
		\begin{equation}
			\frac{dx(t)}{dt} = \mu (1-x(t)) \nonumber
		\end{equation}
	with solution $x(t) = 1- \displaystyle \frac{m}{n} e^{-\mu t}$, given the initial 
	value $x(0) =1- \frac{\displaystyle m}{\displaystyle n} \approx 1$.
	
	We know that when $R_0 > 1$, there is possibility that the epidemic will die out 
	very quickly. Suppose that the epidemic survives and $m=1$. Then, in the early 
	stage of epidemic, we can approximate the process of infection using the birth--
	death process with birth rate $\lambda x(t)$ and death rate $(\gamma + \mu)$. 
	Suppose that $p_i(t)$ denotes the probability $Y_n(t) = i$. Then, the modified 
	forward Kolmogorov equation in (\ref{eq:sir.forward.final}) becomes
		\begin{equation}
			p_i'(t) = (i+1) (\gamma + \mu) p_{i+1}(t) + (i-1) \lambda(t) p_{i-1}(t) - 
				i(\lambda (t)+\gamma + \mu) p_i(t). 
			\label{eq:sir.mod.forward}
		\end{equation}
	From the equation (\ref{eq:sir.forward.final}), the time to extinction of the SIR 
	epidemic must satisfy the differential equation 
		\begin{align}
			p_{\cdot 0 \cdot}'(t) &= p_{\cdot 1 \cdot}(t) \cdot \mu + p_{\cdot 1 \cdot}(t) 
				\cdot \gamma \nonumber \\
			&= (\mu + \gamma) p_{\cdot 1 \cdot}(t) \label{eq:sir.forward.time}
		\end{align}

\begin{figure}[ht]
		\centering
		\subfloat[Parameter $\mu = 0.05$]{%
		\resizebox*{5cm}{!}{\includegraphics{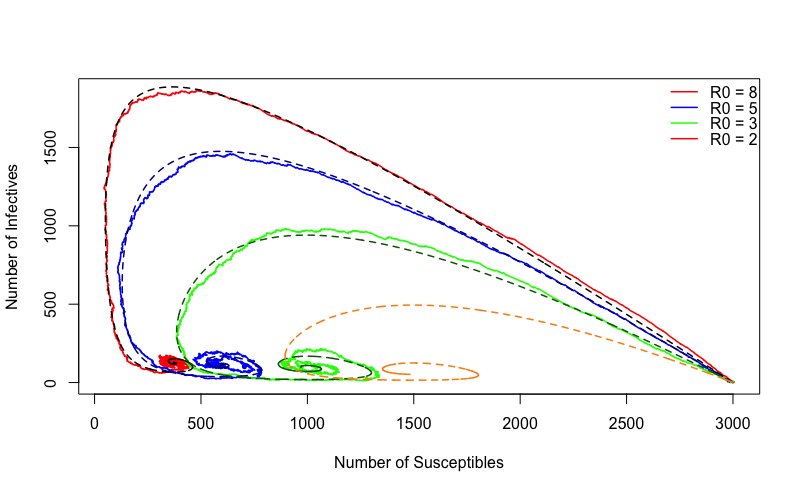}}}
		\hspace{5pt}
 		\label{fig:sir.mu=.05}
		\subfloat[Parameter $\mu = 0.08$]{%
		\resizebox*{5cm}{!}{\includegraphics{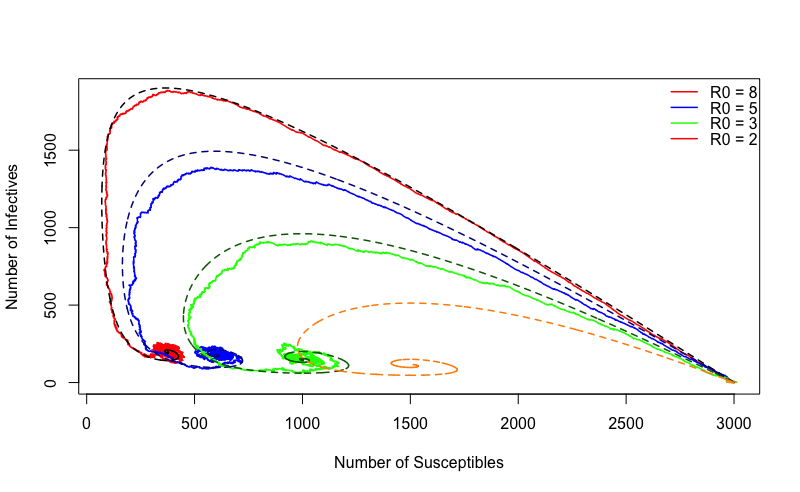}}}
 		\label{fig:sir.mu=.08} \\
		\subfloat[Parameter $\mu = 0.1$]{%
		\resizebox*{5cm}{!}{\includegraphics{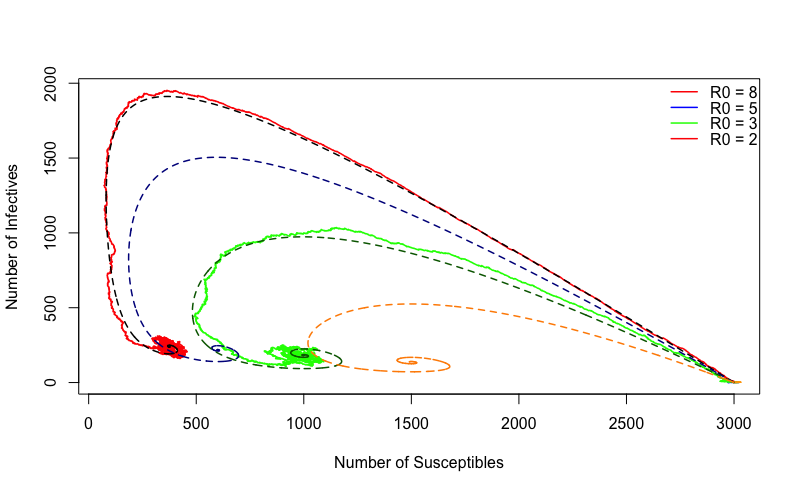}}}
		\hspace{5pt}
		 \label{fig:sir.mu=.1}
		\subfloat[Parameter $\mu = 0.3$]{%
		\resizebox*{5cm}{!}{\includegraphics{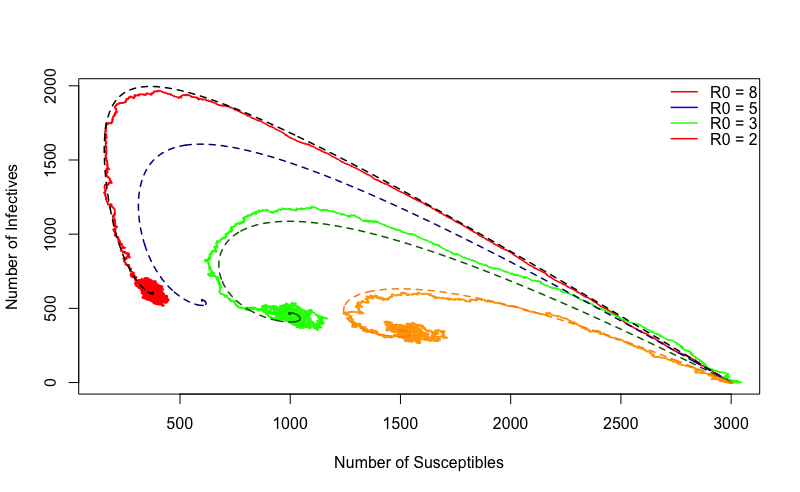}}}
  		\label{fig:sir.mu=.3}
		\caption{SIR with demography with $(n,m,\gamma) = (3000,1,1)$}
		\label{fig:sir.with.various.mu}
	\end{figure}

	The solution of eq. (\ref{eq:sir.mod.forward}) can be seen in Kendall \cite{kendall}
	Now consider Figure \ref{fig:sir.with.various.mu}. When we set $R_0 > 1$, there
	are essentially two behaviours we should notice. First, given the small proportion 
	of initial infectious individuals the infection may die out very quickly as in
	Figure \ref{fig:sir.with.various.mu}. Second, if it survives, then it will enter endemic 
	level at some $t$ as we can see in Figures \ref{fig:sir.r2}, \ref{fig:sir.r4}, and 
	\ref{fig:sir.r8}, where the osilation around the deterministic model becomes 
	damped and the process is now mimicking the deterministic version.
	
\subsection{The Probability of Extinction when $R_0 > 1$}

	The second behaviour can be explained according to the density dependent 
	process and using law of large number \cite{susvitasari}. Now, we need to
	show that there is non--zero probability that the process will die out even
	though $R_0 > 1$.
	
	First, we begin with defining $p_i(t)$ in more detail. The forward Kolmogirov 
	equation is given in eq. (\ref{eq:sir.mod.forward}) where $p_1(0) = 1$ and 
	$p_i(t) = 0$ for $i < 0$. The pgf of $p_i(t)$ is
		\begin{equation}
			\varphi(z,t) = \sum_{k = 0}^\infty p_k(t) z^k \label{eq:pgf.p_i(t)}
		\end{equation}
	and it must satisfy
		\begin{equation}
			\frac{\displaystyle \partial \varphi}{\displaystyle \partial t} = (z-1)(\lambda (t) 
			z - (\gamma + \mu)) \frac{\displaystyle \partial \varphi}{\displaystyle \partial z} 
				\label{eq:lin.par.diff.eq}
		\end{equation}
	with boundary condition $\varphi (z,0) = z$.

	We use the result of Kendall \cite{kendall} that the solution of eq. (\ref{eq:sir.mod.forward})
	is
		\begin{equation}
			p_i(t) = (1-p_0(t)) \cdot (1-\eta_t)\eta_t^{i-1}	\quad \text{for all } i \geq 1
		\end{equation}
	where $\eta_t$ is a function of $t$. Suppose we denote $p_0(t) = \xi$ and $\eta_t
	=\eta \neq 1$, then
		\begin{align}
			\varphi(z,t)  
				&= \xi + \sum_{k=1}^ \infty p_k(t) z^k \nonumber \\
				&= \xi + (1- \xi) (1- \eta) \sum_{k=1} \eta^{k-1} z^k \nonumber \\
				&= \frac{\xi + (1- \xi - \eta)z}{1- \eta z}. \label{eq:pgf.final}
		\end{align}
	Now, by differentiating equation (\ref{eq:pgf.final}) with respect to $t$ and $z$ and 
	then substituting to equation (\ref{eq:lin.par.diff.eq}), yields
		\begin{equation}
			(\eta \xi' - \eta' \xi) + \eta' = \lambda (t) (1 - \eta) (1 - \xi) \label{eq:con.lin.diff.eq}
		\end{equation}
	and recall that $p_0'(t) = \xi' = (\gamma + \mu) p_1(t)$ and using the result from 
	Kendall......... $p_1(t) = (1- \xi)(1- \eta)$. Thus,
		\begin{equation}
			\xi' = (\gamma + \mu) (1- \xi) (1- \eta) \label{eq:xi.prime}
		\end{equation}
	Therefore, by letting $U = 1- \xi$ and $V = 1- \eta$, eq. (\ref{eq:con.lin.diff.eq}) 
	and (\ref{eq:xi.prime}) become
		\begin{subequations}
			\begin{align}
				-U' &= (\gamma + \mu) UV \nonumber \\
				\frac{U'}{U} &= -(\gamma + \mu) V \label{eq:diff.U} \\
				-(1-V) U' + V' (1-U) -V' &= \lambda (t) VU \nonumber \\
				-(1-V) \frac{U'}{U} - V' &= \lambda (t) V \ \nonumber \\
				(\gamma + \mu) (1-V)V - V' &= \lambda (t) V  \nonumber \\
				(\gamma + \mu - \lambda (t)) V - (\gamma + \mu) V^2 &=V'. \label{eq:diff.V}
			\end{align}
		\end{subequations}
	Letting $W = \frac{\displaystyle 1}{\displaystyle V}$ implies that $W' = - 
	\frac{\displaystyle V'}{\displaystyle V^2} = -W^2 V'$, then equation (\ref{eq:diff.V}) 
	becomes
		\begin{align}
			W' + (\gamma + \mu - \lambda (t)) W &= (\gamma + \mu). \label{eq:diff.eq.W}
		\end{align}
	Note that equation (\ref{eq:diff.eq.W}) is the first order differential equation with 
	general solution
		\begin{equation}
			W = \frac{(\gamma + \mu) \displaystyle \int_0^t e^{\rho(u)} du+ c}
				{e^{\rho(t)}} \label{eq:sol.W}
		\end{equation}
	where $\rho(t) = \displaystyle \int_0^t \gamma + \mu - \lambda (u) du$. Now, 
	recall that $p_1(0) = 1$. It implies that $\xi_0 = p_0(0) = 0 = p_i(0) = \eta_0$ 
	for all $i \neq 1$. Consequently, $U_0 = 1 - \xi_0 = 1 = V_0 = 1 - \eta_0$ and 
	$W_0 = \frac{\displaystyle 1}{\displaystyle V_0} = 1$. Therefore, using initial 
	condition at $t=0$ yields
		\begin{equation}
			W = e^{-\rho(t)} \left( 1 + (\gamma + \mu) \int_0^t e^{\rho(u)} du \right).
		\end{equation}
		
	Returning to equation (\ref{eq:diff.U}) and (\ref{eq:diff.V}). Since $V = 
	\frac{\displaystyle 1}{\displaystyle W}$, then it yields $\eta = 1-V = 1 - 
	\frac{\displaystyle 1}{\displaystyle W}$ and $\xi = 1-U = 1- \frac{\displaystyle 
	e^{-\rho(t)}}{\displaystyle W}$. Note that $p_0(t) = \eta$. So,
		\begin{align*}
			p_0(t) &=  1 - \frac{e^{-\rho (t)}}{e^{-\rho(t)} \left( 1 + (\gamma + \mu) 
				\displaystyle \int_0^t e^{\rho(u)} du \right)} \nonumber \\
			&=  1 - \frac{1}{ \left( 1 + (\gamma + \mu) \displaystyle \int_0^t e^{\rho(u)} 
			du \right)}
		\end{align*}
	where $\rho(t) = \displaystyle \int_0^t (\gamma + \mu) - \lambda(u) du$.
	
	Note that if we let $R_0 = \frac{\displaystyle \lambda(t)}{\displaystyle \gamma 
	+ \mu} >1$, then equation (\ref{eq:sir.final.prob.ext}) lays in $(0,1)$.
	This explains mathematically the extinction in Figure \ref{fig:sir.with.various.mu} 
	and also supports the branching process theory. Otherwise, by letting $R_0 \leq 1$, 
	and if we let $t \rightarrow \infty$, then $\left( 1 + ( \gamma + \mu) \displaystyle 
	\int_0^t e^{\rho(t)} du \right)  \rightarrow \infty$. Consequently, $p_0(t) \rightarrow 1$ 
	as $t \rightarrow \infty$. This result is not surprising since we have already known 
	that when $R_0 \leq 1$, the stable stage of the epidemic process is attained in the 
	disease-free stage.

\section{Conclusion}

According to the threshold theorem, epidemic can only occur if the initial number of susceptibles is larger than some critical value, which depends on the parameters under consideration (Ball, 1983). Usually, it is expressed in terms of epidemic reproductive ratio number, $R_0$. This quantity is defined as the expected number of contacts made by a typical infective to susceptibles in the population.

We have successfully showed that both deterministic and stochastic models performed similar results when $R_0 \leq 1$. That is, the disease-free stage in the epidemic. But when $R_0 > 1$, the deterministic and stochastic approaches had different interpretations. In the deterministic models, both the SIS and SIR models showed an outbreak of the disease and after some time $t$, the disease persisted and reached endemic-equilibrium stage. The stochastic models, on the other hands, had different interpretations. If we let the population size be sufficiently large, the epidemic might die out or survive. There were essentially two stages to this model. First, the infection might die out in the first cycle. If it did, then it would happen very quickly, just like the branching process theory described. Second, if it survived the first cycle, the outbreak was likely to occur, but after some time $t$, it would reach equilibrium just like the deterministic version. In fact, the stochastic models would mimic the deterministic's paths and be scattered randomly around their equilibrium point.

\end{document}